# Superconducting nanowire photon number resolving detector at telecom wavelength

Aleksander Divochiy[1], Francesco Marsili[2,*,†], David Bitauld[2,†], Alessandro Gaggero[3], Roberto Leoni[3], Francesco Mattioli[3], Alexander Korneev[1], Vitaliy Seleznev[1], Nataliya Kaurova[1], Olga Minaeva[1], Gregory Gol'tsman[1], Konstantinos G. Lagoudakis[2], Moushab Benkhaoul[4], Francis Lévy[4], Andrea Fiore[2,†]

[*]**e-mail: francesco.marsili@epfl.ch**
[1]**Moscow State Pedagogical University (MSPU), Department of Physics, 119992 Moscow, Russian Federation**
[2]**Ecole Polytechnique Fédérale de Lausanne (EPFL), Institute of Photonics and Quantum Electronics (IPEQ), Station 3, CH-1015 Lausanne, Switzerland**
[3]**Istituto di Fotonica e Nanotecnologie (IFN), CNR, via Cineto Romano 42, 00156 Roma, Italy**
[4]**Ecole Polytechnique Fédérale de Lausanne (EPFL), Institute of Complex Matter Physics (IPMC), Station 3, CH-1015 Lausanne, Switzerland**

[†]**Present address: Eindhoven University of Technology, P.O. Box 513, NL-5600MB Eindhoven, The Netherlands**

The optical-to-electrical conversion, which is the basis of optical detectors, can be linear or nonlinear. When high sensitivities are needed single-photon detectors (SPDs) are used, which operate in a strongly nonlinear mode, their response being independent of the photon number. Nevertheless, photon-number resolving (PNR) detectors are needed, particularly in quantum optics, where *n*-photon states are routinely produced. In quantum communication, the PNR functionality is key to many protocols for establishing, swapping and measuring entanglement, and can be used to detect photon-number-splitting attacks. A linear detector with single-photon sensitivity can also be used for measuring a temporal waveform at extremely low light levels, e.g. in long-distance optical communications, fluorescence spectroscopy, optical time-domain reflectometry. We demonstrate here a PNR detector based on parallel superconducting nanowires and capable of counting up to 4 photons at telecommunication wavelengths, with ultralow dark count rate and high counting frequency.

Among the approaches proposed so far to PNR detection, detectors based on charge-integration or field-effect transistors[1-3] are affected by long integration times, leading to bandwidths <1 MHz. Transition edge sensors (TES[4]) operate at 100 mK and show long response times (several microseconds). Approaches based on photomultipliers (PMTs)[5] and avalanche diodes (APDs), such as the visible light photon counter (VLPC)[6], 2D arrays of APDs[7,8] and time-multiplexed detectors[9] are not sensitive or they are plagued by high dark count rate and long dead times in the telecommunication spectral windows. Arrays of SPDs additionally involve complex read-out schemes[8] or separate contacts, amplification and discrimination[10]. The parallel nanowire detector (PND) presented here significantly outperforms these approaches in terms of simplicity, sensitivity, speed, and multiplication noise.

The basic structure of the PND is the parallel connection of $N$ superconducting nanowires, each connected in series to a resistor $R_0$ (Fig. 1). The detecting element is a few nm-thick, ≈100 nm-wide NbN wire folded in a meander pattern. Each branch acts as a superconducting single photon detector[11] (SSPD). If a superconducting nanowire is biased close to its critical current, the absorption of a photon causes the formation of a normal barrier across its cross section and the bias current is pushed to the external circuit. In the parallel configuration proposed here, the currents from different wires can sum up on the external load, producing an output voltage pulse proportional to the number of photons. The time evolution of the device after photon absorption can be simulated using the equivalent circuit of Fig. 1b. Let $I^{(i)}=I_B=V_B/R_0$ be the current flowing through each section when the device is biased with a voltage source $V_B$. If a photon reaches the $i$-th nanowire, it will cause the superconducting-normal transition with a probability $\eta_i = \eta\left(I_B / I_C^{(i)}\right)$, where $\eta$ is the current-dependent quantum efficiency and $I_C^{(i)}$ is the critical current of the nanowire[11] (the nanowires may have different critical currents, being differently constricted[12]). Because of the sudden increase in the resistance of the nanowire, its current is then redistributed between the other N-1 branches and the input resistance of the amplifier $R_A$=50 Ω, (fig. 2a and 2b). The peak value of the output current is then $I_{out}(1)$~$(I_B-(N-1)·\delta I_{lk})$, $\delta I_{lk}$ being the leakage current drained by each of the still superconducting nanowire. If $m$ ($m\leq N$) photons are simultaneously absorbed in $m$ distinct nanowires, the currents from different branches sum up on $R_A$, producing a current pulse of height $I_{out}(m)$~$(m·I_B-m(N-m)·\delta I_{lk})$. The device shows PNR capability if the remaining wires do not shunt the load, i.e. if $\delta I_{lk} \ll I_B$. The leakage current is also undesirable because it lowers the signal available for amplification and temporary increases the $I^{(i)}$'s, eventually driving other nanowires normal. The leakage current can be reduced by engineering the dimensions of the nanowire (thus its kinetic inductance[13]) and of the bias resistor to maximize the wire impedance $Z_B$ while keeping a stable bias condition and a short response time. A design without bias resistors is also possible. It simplifies the fabrication process, but, as $Z_B$ is lower, $\delta I_{lk}$ significantly limits the maximum bias current allowed for the stable operation of the device and then its quantum efficiency.

PNDs were fabricated on ultrathin NbN films (4nm) on MgO[14] and R-plane sapphire[15] using electron beam lithography (EBL) and reactive ion etching (see methods). Detector size ranges from 5x5 μm² to 10x10 μm² with the number of parallel branches varying from 4 to 14. The nanowires are 100 to 120 nm wide and the fill factor of the meander is 40 to 60%. The length of each nanowire ranges from 25 to 100 μm. Designs with and without the integrated bias resistors were tested. The scanning-electron microscope image of a PND with six parallel wires (6-PND) and series resistors fabricated on MgO is shown in Fig. 1a.

The photoresponse of a 10x10 μm$^2$ 4-PND probed with light at 1.3 μm was recorded by a sampling oscilloscope (inset Fig. 2c). All four possible amplitudes can be observed. The pulses show a full width at half maximum (FWHM) as low as 660ps. PNDs showed counting performance when probed with light at 26 and 80 MHz repetition rate (fig. 2c and inset of fig. 4a, respectively), outperforming any existing PNR detector at telecom wavelength by three orders of magnitude. Indeed, PNDs have by design a reduced recovery time (by a factor $N^2$) even compared to traditional SSPDs, due to the reduced kinetic inductance[16].

In order to prove that the PND response is proportional to the photon number, we measured the photocount statistics. The photon number probability distribution measured with a PNR detector $Q(n)$ is related to the incoming distribution $S(m)$ by the relation: $Q(n) = \sum_{m \geq n} P(n|m) \cdot S(m)$, where $P(n|m)$ is the probability that $n$ photons are detected when $m$ are sent to the device. The PND is easily probed with an attenuated laser, which has a photon number distribution close to a Poissonian $S(m)=\mu^m \cdot \exp(-\mu)/m!$ ($\mu$: mean photon number). In the regime $\mu \ll 1$, $S(m) \sim \mu^m/m!$, and for $\mu$ low enough the measured distribution can be written as:

$$Q(n) \sim P(n|n) \cdot S(n) \propto \mu^n / n! \qquad (1)$$

Consequently, the probability $Q(1)$ of detecting one photon is proportional to $\mu$, $Q(2)$ is proportional to $\mu^2$, and so on. A 5-PND was tested with the coherent light from a 850 nm GaAs pulsed laser. In fig. 3a the detection probabilities relative to the one, two and three-photon absorption events are plotted for $\mu$ varying over two orders of magnitude. As predicted by (1), $Q(n) \propto \mu^n$, which demonstrates the PNR functionality. The one-photon quantum efficiency $\eta$ at 1.3 μm and dark-counts rate DK were measured as a function of bias current (fig. 3b). The lowest DK value measured was 0.15 Hz for $\eta=2\%$ (yielding a noise equivalent power[17] $NEP=5.6 \times 10^{-18}$ W/Hz$^{1/2}$), limited only by the room temperature background radiation coupling to the PND.

The typical application of a PNR detector is the reconstruction of an unknown photon number distribution $S(m)$, which can be recovered given $Q(n)$ and the matrix of the conditional probabilities[18]. Let $P^N = \left[ P^N_{n,m} \right]$, with $P^N_{n,m} = P^N(n|m)$, be the matrix of the conditional probabilities for an $N$-PND. Considering equations (M.1) to (M.3) in methods, it is clear that $P^N$ can be calculated if the vector of the $N$ different quantum efficiencies $\bar{\eta} = [\eta_i]$ is known. $\bar{\eta}$ can be determined fitting the $Q(n)$ measured when probing the device with a light whose $S(m)$ is known. A 5-PND was tested with the coherent emission from a Ti:Sapphire laser. To determine $Q(n)$, histograms of the photoresponse voltage peak $V_{pk}$ were built for values of $\mu$ ranging from ~1 to ~100 (fig. 4). The experimental probability distribution $Q(n)$ measured for different $\mu$ was then fitted to the one predicted by the model (see methods) using the vector $\bar{\eta}$ as free parameter (Fig. 5). The photocount statistics of six levels is well fitted over almost two orders of magnitude of mean photon number, confirming the validity of the model. Additionally, the fitted efficiencies (inset of Fig. 5) are rather uniform, indicating a high-quality fabrication process.

Several effects may limit the counting capability $M_{max}$ of a PNR detector. One is the quantum efficiency. From equation (M.2) in methods, assuming the detector saturation is negligible ($n \ll N$) and that all the branches are equal ($\eta_i=\eta$), the probability $Q(n)$ of detecting $n$ photons is proportional to $\eta^n$. In the PND tested $\eta \sim 2\%$ at 1.3 $\mu$m, which we attribute to unoptimised film thickness and device design. The $\eta$ of SPDs based on the same detection mechanism can be increased up to $\sim 60\%$[19], and could potentially exceed 90% using optimized optical cavities. The second limitation is the electrical noise. Pulse height discrimination can be performed as long as the noise remains lower than the one-photon signal amplitude. In most PNR detectors[1-3,5-7,20] the noise increases with the detection level due to multiplication noise, which limits the maximum number of resolvable photons. In contrast, no multiplication noise is seen in PNDs, as the width of the histogram peaks is independent of the number of detected photons $n$ (fig. 4). Indeed we expect an excess noise factor $F$[21] close to unity in the PND, since the noisy quasi-particle multiplication process[22] causes a fluctuation only in the hotspot resistance $R_{hs}$ ($\sim 1$ k$\Omega$) and not in the output current, the latter being determined by the partition with the much lower load resistance $R_A$. A third limitation to $M_{max}$ in PNDs is the leakage current $\delta I_{lk}$, which restricts the number of parallel wires. However, this issue can be overcome by switching from voltage- to current read-out (e.g. using a transimpedance amplifier), thus decreasing the load impedance.

In conclusion, a new PNR detector, the Parallel Nanowire Detector, has been demonstrated, which significantly outperforms existing approaches in terms of sensitivity, speed and multiplication noise in the telecommunication wavelength range. In particular, it provides a repetition rate (80 MHz) three orders of magnitude larger than any existing detector at telecom wavelength[1,4,8], and a sensitivity (NEP=5.6x10$^{-18}$ W/Hz$^{1/2}$) one-two orders of magnitude better (with the exception of transition-edge sensors[4], which require a much lower operating temperature). The high repetition rate and high sensitivity make it already suitable – for the first time – for replacing correlation set-ups in quantum optics experiments at telecommunication wavelengths. By increasing the efficiency, the performance needed for the single-shot measurement of photon number, as needed in many quantum communication and computing protocols, can be reached. Finally, increasing the maximum photon number to 20-30 photons, the PND could be used as an "analog" detector with single-photon sensitivity, bridging the gap between conventional and single-photon detectors.

## METHODS

### FABRICATION

NbN films were grown on sapphire or MgO substrates by reactive magnetron sputtering in an argon–nitrogen gas mixture. Using an optimized sputtering technique, our NbN samples exhibited a superconducting transition temperature of $T_c$ =10.5 K for 40-Å-thick films. The superconducting transition width was equal to $\Delta T_c$ = 0.3 K.

For the devices on MgO, the three nanolithography steps needed to fabricate the structure have been carried out by using an electron beam lithography (EBL) system equipped with a field emission gun (acceleration voltage 100 kV). In the first step pads and alignment markers (60 nm Au on 10nm Ti) are fabricated by lift off via a Polymethyl Methacrylate (PMMA, a positive tone electronic resist) stencil mask. In the second step, an hydrogen silsesquioxane (HSQ, a negative tone electronic resist) mask is defined reproducing the meander pattern. All the unwanted material, i.e. the material not covered by the HSQ mask and the Ti/Au film, is removed by using a fluorine based reactive ion etching (RIE). Finally, with the third step the bias resistors (85nm AuPd alloy, 50%-each in weight) aligned with the two previous layers are fabricated by lift off via a PMMA stencil mask.

Details on the fabrication process of the devices on sapphire can be found in[16].

### MEASUREMENT SETUP

Electro-optical characterizations have been performed in a cryogenic probe station with an optical window and in cryogenic dipsticks. Bias current was supplied by a low noise voltage source in series with a bias resistor (through the DC port of a bias-T). The AC port of the bias-T was connected through to high-bandwidth, low-noise amplifiers. The amplified signal was fed either to a 1 GHz bandwidth single-shot oscilloscope, to a 40 GHz bandwidth sampling oscilloscope, or to a 150 MHz counter. The optical input was provided by a fiber-pigtailed, gain-switched laser diode at 1.3 μm wavelength, a mode-lock Ti:sapphire laser at 700 nm wavelength, or an 850 nm GaAs pulsed laser.

In the cryogenic probe station (Janis) the devices were tested at a temperature T=5 K. Electrical contact was realized by a cooled 50 Ω microwave probe attached to a micromanipulator, and connected by a coaxial line to the room-temperature circuitry. The light was fed to the PNDs through a single-mode optical fiber coupled with a long working distance objective, allowing the illumination of a single detector.

In the cryogenic dipsticks the devices were tested at 4.2 K or 2 K. The light was sent through a single-mode optical fiber either put in direct contact and carefully aligned with the active area of a single device or coupled with a short focal length lens, placed far from the plane of the chip in order to ensure uniform illumination. The number of incident photons per device area was estimated with an error of 5 %.

### MODELLING

Assuming that the illumination of the device is uniform, the parallel connection of N nanowires can be considered equivalent to a balanced lossless N port beam splitter, every channel terminating with a single photon detector (SPD). Each incoming photon is then equally likely to reach one of the N SPDs (with a probability 1/N). Each SPD can detect a photon with a probability $\eta_i$ (i=1,..,N) different from all the others, and gives the same response for any number (m≥1) of photons detected. The signals from all the SPDs are then summed up to obtain the output. Following[23], two classes of terms in $P^N$ be calculated directly, the others being derived from these by a recursion relation. These terms are the probabilities $P^N_{m,m}$ that all the $m \leq N$ photons sent are detected and $P^N_{0,m}$ that no photons are detected when m are sent. In the case of zero detections, $P^N_{0,m}$ is given by:

$$P_{0,m}^{N} = \sum_{i_1=1,\ldots,i_m=1}^{N} \left[ \frac{1-\eta_{i_1}}{N} \cdot \ldots \cdot \frac{1-\eta_{i_m}}{N} \right] \qquad (M.1)$$

which assumes that a photon incident in the i-th nanowire fails to be detected with an independent probability of $(1-\eta_i)$. For the case where all the photons are detected, since m photons must reach m distinct nanowires:

$$P_{m,m}^{N} = \sum_{\substack{i_1=1,\ldots,i_m=1 \\ i_p \neq i_q \text{ for } p \neq q}}^{N} \left[ \frac{\eta_{i_1}}{N} \cdot \ldots \cdot \frac{\eta_{i_m}}{N} \right] \quad \text{for } m \leq N \qquad (M.2)$$

The recursion relation for $P_{nm}^{N}$ is:

$$P_{n,m}^{N} = P_{n,m-1}^{N} \left[ \frac{n}{N} + \frac{n!}{N!} \cdot \sum_{\substack{i_1=1,\ldots,i_{N-n}=1 \\ i_p \neq i_q \text{ for } p \neq q}}^{N} \left( \frac{1-\eta_{i_1}}{N} + \ldots + \frac{1-\eta_{i_{N-n}}}{N} \right) \right] + $$
$$+ P_{n-1,m-1}^{N} \left[ \frac{(n-1)!}{N!} \cdot \sum_{\substack{i_1=1,\ldots,i_{N-(n-1)}=1 \\ i_p \neq i_q \text{ for } p \neq q}}^{N} \left( \frac{1-\eta_{i_1}}{N} + \ldots + \frac{1-\eta_{i_{N-(n-1)}}}{N} \right) \right] \qquad (M.3)$$

The first term on the right-hand side is the probability that *n* photons are detected when *m*-1 are sent, times by the probability that the *m*-th photon reaches one of the *n* nanowires already occupied or that it fails to be detected reaching one of the *N-n* unoccupied nanowires. The second term is the probability that *n*-1 photons are detected when *m*-1 are sent times by the probability that the *m*-th photon reaches one of the *N-n*+1 unoccupied nanowires and it is detected. In the limit $\eta_i = \eta$ for *i*=1,...,*N*, the recursion relation agrees with that given in[23].

## ACKNOWLEDGEMENTS

This work was supported by: Swiss National Foundation through the "Professeur boursier", NCCR Quantum Photonics program, FP6 STREP "SINPHONIA" (contract number NMP4-CT-2005-16433), IP "QAP" (contract number 15848), the grant "Non-equilibrium processes after IR photon absorption in thin-film superconducting nanostructures" of Russian Agency on education and the grant 02.445.11.7434 of Russian Ministry of education and science for support of leading scientific schools. The authors thank B. Deveaud-Plédran, B. Dwir and H. Jotterand for useful discussion and technical support and the Interdisciplinary Centre for Electron Microscopy (CIME) for supplying TEM and SEM facilities.

## COMPETING FINANCIAL INTERESTS STATEMENT

The authors declare that they have no competing financial interests.

21  McIntyre, R.J. Multiplication noise in uniform avalanche diodes. *IEEE Trans. Electron Devices* 13 (1), 164 - 168 (1966).
22  Semenov, A. D., Gol'tsman, G. N., & Korneev, A. A. Quantum detection by current carrying superconducting film. *Physica C* 351 (4), 349-356 (2001).
23  Fitch, M. J., Jacobs, B. C., Pittman, T. B., & Franson, J. D. Photon-number resolution using time-multiplexed single-photon detectors. *Phys. Rev. A* 68 (4 B) (2003).

## FIGURE LEGENDS

**Figure 1 The Parallel Nanowire Detector (PND). a,** Scanning electron microscope (SEM) image of an PND with $N$=6 (6-PND) fabricated on a 4nm thick NbN film on MgO. The nanowire width is $w$=100 nm, the meander fill factor is $f$=40%. The detector size is 10x10 μm$^2$. The active nanowires (in color) are connected in series with Au-Pd bias resistors (in blue). The floating meanders at the four corners of the PND pixel correct for the proximity effect. The devices are contacted through 70nm thick Au-Ti pads, patterned as a 50 Ω coplanar transmission line. **b,** Equivalent circuit of the 6-PND. The superconducting nanowire is modeled as the series of an inductance $L_{kin}$ accounting for its kinetic inductance, a switch $S_L$ which opens on the hotspot resistance $R_{hs}$~1 kΩ, simulating the absorption of a photon. Each branch is the series of a nanowire and of the bias resistor $R_0$. The device is connected through a bias T to the bias voltage source $V_B$ and to the radio-frequency amplifier with input resistance $R_A$=50 Ω.

**Figure 2 The PND photoresponse. a,** Simulation result for the time evolution during photodetection of the currents $I^{(i)}$ ($i$=1,...,4) flowing through the branches of a 4-PND with $\eta_i$=1. The curves are shifted vertically for clarity. In the simulation five light pulses (containing one, three, four, two and three photons) are sent to the device, with a repetition frequency of 26MHz. The photons in each pulse are never absorbed on the same nanowire, so the number of incident photons is equal to the number of branches making a transition. The sections absorbing the photon experience a large drop in their current, the others experience an increment which is a multiple of $\delta I_{lk}$. **b,** Simulation result for the current $I_{out}$ flowing through $R_A$ (continuous curve), and number of incident photons in each light pulse (□). The curves relative to $I_{out}$ and the $I^{(i)}$ are plotted in the same units. **c,** Single-shot oscilloscope trace during photodetection by a 10x10 μm$^2$ 4-PND (with integrated bias resistors). The device was probed in the cryogenic probe station under illumination with 1.3 μm, 100ps-long pulses from a laser diode, at a repetition rate of 26MHz. The five successive response pulses have clearly four discrete amplitudes. Inset: Photoresponse transients taken with a 40 GHz sampling oscilloscope (□). The orange solid curves are guides to the eyes.

**Figure 3 *n*-photon detection probability vs incoming mean photon number.** The curves were built from the photoresponse of a 5-PND (with integrated bias resistors). The device was tested under uniform illumination in a cryogenic dipstick dipped in a liquid He bath at 2.2 K. **a,** Detection probabilities relative to the one (□), two (□) and three-photon (□) absorption events as a function of the mean photon number per pulse $\mu$, varying from 0.15 to 40. The light pulses at 0.85μm form the GaAs pulsed laser were 30 ps wide and the repetition rate was 100kHz. The average input photon number $\mu$ was set with a variable fiber-based optical attenuator. The photoresponse from the device was sent to the 150 MHz counter. The light intensities and the thresholds of the counter discriminator were chosen so that for any $Q(n)$ the counts corresponding to an *n*-photon absorption event were significantly higher than the counts corresponding to the absorption of more than *n* photons. The fittings clearly show that $Q(1) \propto \mu$, $Q(\mu,2) \propto \mu^2$ and $Q(\mu,3) \propto \mu^3$, which demonstrates the capability of the detector to resolve one, two and three photons simultaneously absorbed.. **b,** One-photon quantum efficiency η and dark-counts rate DK as a function of bias current. The device was probed with a 1.3 μm, laser diode.

**Figure 4 Histograms of the photoresponse voltage peak.** Histograms were built by sampling the photoresponse of an 8.6x8 μm$^2$ 5-PND (with no integrated bias resistors). The device was tested under uniform illumination in a cryogenic dipstick dipped in a liquid He bath at 4.2 K. The light pulses at 700nm form a Ti:sapphire laser were 40ps wide (after the propagation in the optical fiber) and the repetition rate was 80MHz. The average input photon number per pulse $\mu$ was set with a free space variable optical attenuator. The signal from the device was sent to the 1 GHz oscilloscope, which was triggered by the synchronization generated by the laser unit. The photoresponse was sampled for a gate time of 5ps, making the effect of dark counts negligible. Increasing $\mu$, e.g. form 7.7 (a) to 97.6 (b), the shape of the histograms changes as the probability to observe higher response amplitudes increases. The solid lines are the experimental histograms. The dashed lines represent the fitted gaussian distribution of each possible pulse level. Inset: Single-shot oscilloscope trace during photodetection at 80MHz repetition rate.

**Figure 5 Experimental probability distribution vs mean photon number.** The experimental discrete probability distribution $Q(n)$ (*) was estimated from the continuous probability density $q(V_{pk})$ (fig. 4) fitting the histograms to the sum of 6 gaussian distributions (corresponding to the five possible pulse levels plus the zero level) and calculating their area. The 5-PND was probed with several mean photon numbers $\mu$: 1.5, 2.8, 4.3, 5.3, 7.7, 12.5, 15.9, 26.9, 33.6, 64.9. The experimental values for $Q(n)$ were then fitted (●) using a genetic algorithm to recover the vector of quantum efficiencies $\overline{\eta}$. Inset: Reconstructed quantum efficiencies of each section of the 5-PND. The standard deviation is 20% of the mean value, proof of the excellent uniformity of the device.

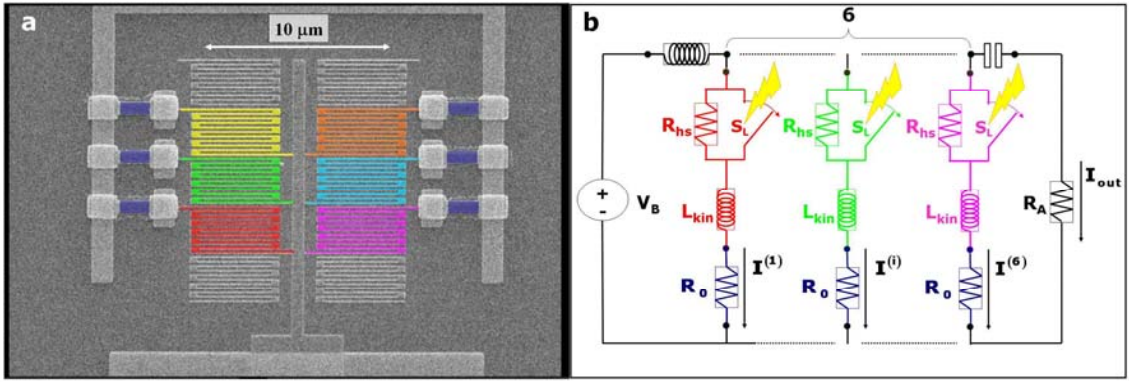

**Figure 1**

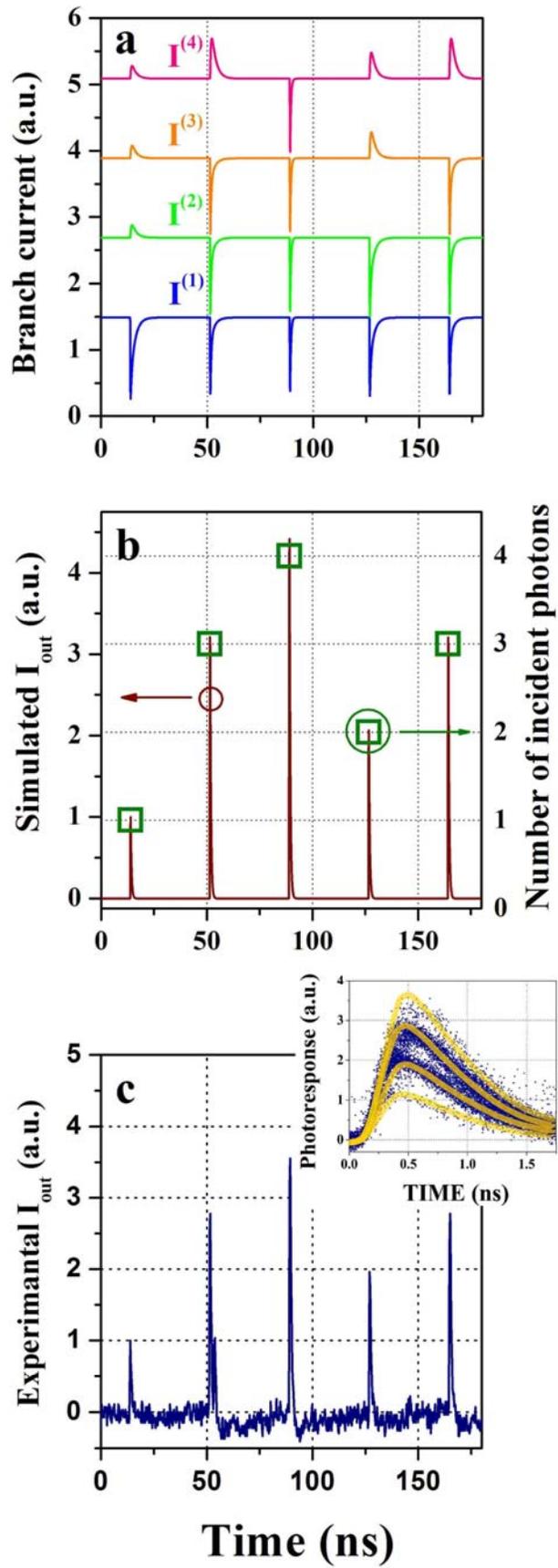

Figure 2

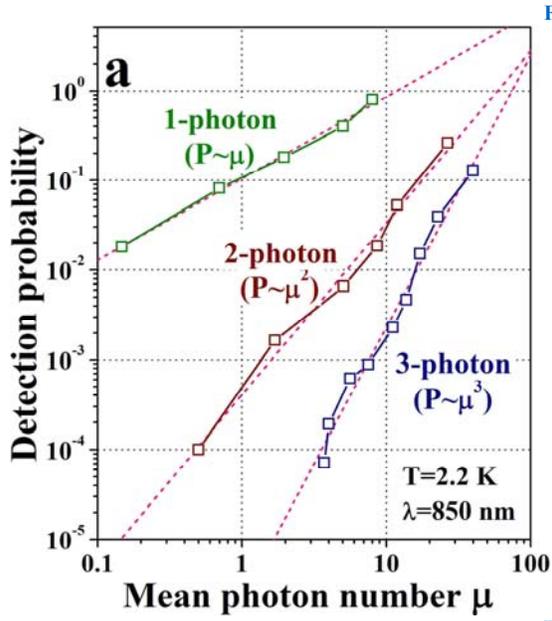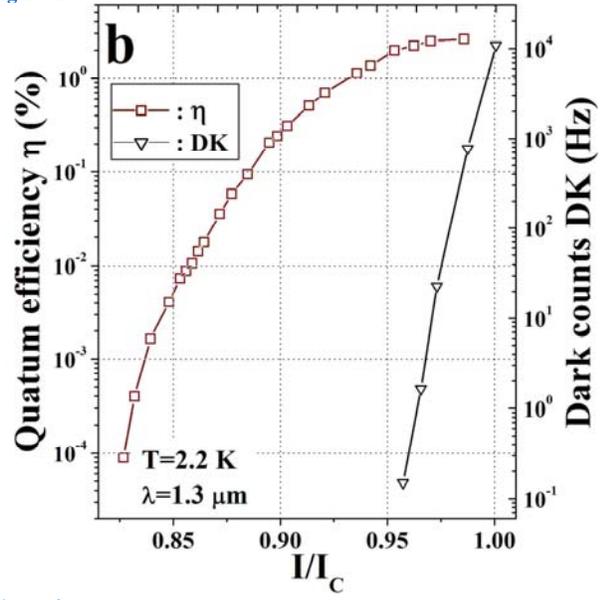

Figure 3

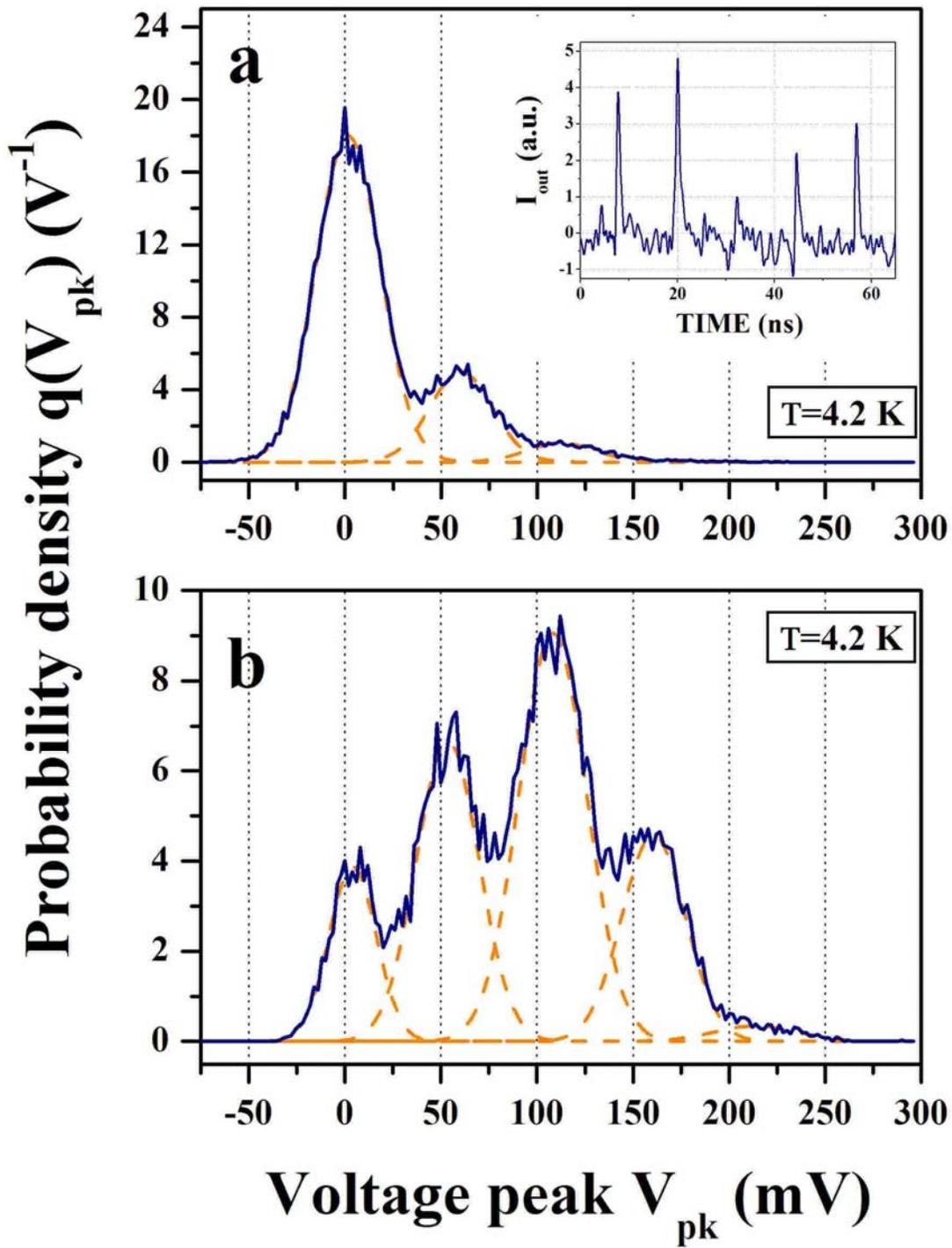

Figure 4

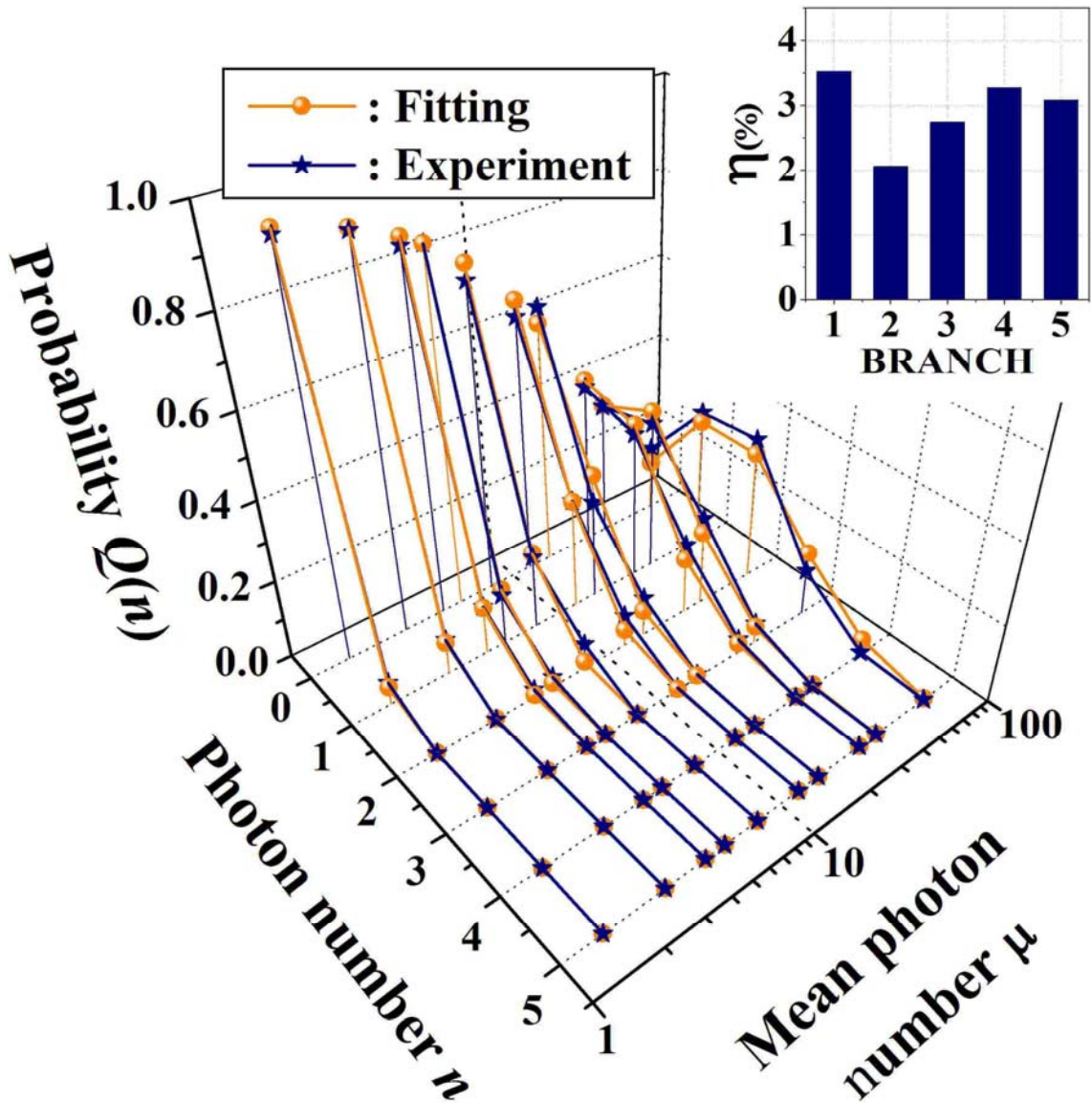

**Figure 4**